\input phyzzx\def\e{\adveq\eqno{\rm (\chapterlabel.\the\equanumber)}}
\def\mysec#1{\equanumber=0\chapter{#1}}
\def\adveq{\global\advance\equanumber by 1}
\def\myeq{{\rm \chapterlabel.\the\equanumber}}
\def\rarrow{\rightarrow}

\def\semidirect{\mathrel{\raise0.04cm\hbox{${\scriptscriptstyle |\!}$
\hskip-0.175cm}\times}}

%\define\semidirect{\propto}

\def\ref#1{$^{[#1]}$}

\def\r#1{$[\rm#1]$} 
\def\twidle{\tilde}

\def\Tr{\mathop{\rm Tr}\limits}
\def\e{\adveq\eqno{\rm (\chapterlabel.\the\equanumber)}}
\def\mysec#1{\equanumber=0\chapter{#1}}
\def\adveq{\global\advance\equanumber by 1}
\def\myeq{{\rm \chapterlabel.\the\equanumber}}
\def\rarrow{\rightarrow}

\def\semidirect{\mathrel{\raise0.04cm\hbox{${\scriptscriptstyle |\!}$
\hskip-0.175cm}\times}}

%\define\semidirect{\propto}

\def\ref#1{$^{[#1]}$}

\def\r#1{$[\rm#1]$} 
\def\twidle{\tilde}

\def\Tr{\mathop{\rm Tr}\limits}

\def\half{{1\over2}}

\date{December, 2021}
\date{December, 2021}
\titlepage
\title{The Crossing Multiplier for Solvable Lattice Models}
\author{Vladimir Belavin$^a$, Doron Gepner$^b$ and J. Ramos Cabezas$^a$}
\vskip20pt
\line{\it \hfill $^a$ Physics Department, Ariel University, Ariel 40700, Israel,\hfill}
\line{\it\hfill $^b$  Department of Particle Physics and Astrophysics, Weizmann Institute,\hfill}
 \line{\it\hfill Rehovot 76100,  Israel\hfill} 
 
 \abstract
 We study the large class of solvable lattice models, based on the data of conformal field theory. 
 These models are constructed from any conformal field theory.
 We consider the lattice models based on affine algebras described by Jimbo et al., for the algebras
$ABCD$ and by Kuniba et al. for $G_2$. We find a general formula for the crossing multipliers
of these models. It is shown that these crossing multipliers are also given by the principally specialized
characters of the model in question. Therefore we conjecture that the crossing multipliers in this large
class of solvable interaction round the face lattice models are given by the characters of the conformal field theory on which they are based. We use this result to study the local state probabilities of these
models and show that they are given by the branching rule, in regime III.

 \endpage

\mysec{Introduction}

Solvable lattice models in two dimensions are a fruitful ground to test phase transitions, universality
and two dimensional condensed matter systems. For a review see 
\REF\Baxter{R.J. Baxter, ``Exactly solved models in statistical mechanics" (1982).}
\r\Baxter.
\REF\Found{D. Gepner, ``Foundation of rational quantum field Theory I, hep-th/9211100}
An approach to solvable Interaction Round the Face (IRF) 
lattice models was presented where the lattice model itself is built out of the data of some conformal field
theory and two primary fields in this conformal field theory
\r\Found. On each vertex of the model sits a  primary field
in the theory and the admissibility condition is given by the fusion rules. For an explanation of  
conformal field theory (CFT)  
\REF\BPZ{A.A. Belavin, A.M. Polyakov and A.B. Zamolodchikov, ``Infinite conformal symmetry in
two--dimensional quantum field theory", Nucl. Phys. B 241 (1984) 333.}
\REF\Francesco{P. Di Francesco, P. Mathieu and D. Senechal, ``Conformal field theory",
Springer Verlag, New York (1997).}
see \r\BPZ, and the review \r\Francesco. The aim of this paper is to treat the local state probabilities
of such IRF models. This would further the understanding of such models.

The solution is based on the Baxterization of the braiding matrix and
is a trigonometric solution to the Yang Baxter equation (YBE) \r\Found. Our aim here is to extend this 
solution to the elliptic (thermalized) case.  
\REF\Free{D. Gepner,  ``On the free energy of solvable lattice models", Nucl. Phys. B 971 (2021) 115532.}
The inversion relations
of the general elliptical IRF model were conjectured and based on this the free energy was calculated
in the four main regimes \r\Free. Our aim here is to enlarge this result to the local state probabilities.
In ref. \r\Found, the crossing multiplier for the trigonometric solution was proposed
to be given by the modular matrix. We extend this conjecture to the elliptic case by proposing
that the crossing multipliers are given, in general, by the characters. The proof of this for WZW models
is described in the appendix.

In ref. \r\Free\ it was conjectured that in regime III the fixed point field theory is given by the coset
$G/\cal O$, where $G$ is some CFT and $\cal O$ is the original CFT used to build the model.
We find support for this conjecture by calculating the local state probabilities of these models.
We find that it is given by the branching function of the coset multiplied by the character,
up to normalization, which is given by the branching rule.

\mysec{The IRF models and their crossing relations} 

We define the IRF lattice models  
based on some rational conformal field theory (RCFT), $\cal O$, and a pair of primary 
fields in this RCFT denoted by $h$ and $v$ 
\r\Found.
The model is denoted accordingly by IRF$({\cal O},h,v)$.
For simplicity, we assume that $h=v$.
We define the models on a square lattice, where on each vertex sits some primary field. We assume that
the face Boltzmann weight vanishes unless the admissibility condition is obeyed, which is,
$$
f_{a, h}^b>0,\quad  f_{c ,h }^d>0,\quad f_{a,h}^c>0,\quad f_{b,h}^d>0, \e
$$
where $a,b,c,d$ are the four primary fields sitting on a face and $f_{x,y}^z$
is the fusion coefficient in the RCFT $\cal O$.
For an explanation of these notions see e.g. 
\r\Francesco.
The partition function of the model is 
\def\om#1#2#3#4#5{\omega\left( \matrix{#1 & #2 \cr #3 & #4 \cr}\bigg | #5 \right)}
$$
Z=\sum_{\rm {configurations}}\prod_{\rm {faces}} \om a b c d u.\e$$ 
where $\omega$ is the Boltzmann weight and $u$ is the spectral parameter.

We wish to define the Boltzmann weight $\omega$ in such a way that the model will be solvable.
Namely, that the transfer matrices will commute for different spectral parameters. This is 
guaranteed by the Yang--Baxter equation (YBE), see, e.g., 
\r\Baxter.
It is simpler to define this equation in operator form.
For this, we define the operator,
$$
<a_1,a_2,\ldots, a_n| R_i(u)  |a_1^\prime,a_2^\prime,\ldots,a_n^\prime>=
\om {a_{i-1}} {a_i}  {a_i^\prime}  {a_{i+1}} u
\prod_{m=1\atop m\neq i}^n\delta_{a_m,a_m^\prime}, \e$$
Then the YBE assumes the form,
$$
R_{i+1}(u) R_i(u+v) R_{i+1}(v)=R_i(v) R_{i+1}(u+v) R_i(u),
\e$$

A trigonometric solution of the YBE was conjectured  for any RCFT
$\cal O$ and for any $h$ and $v$, provided that the fusion coefficients of $h$ and $v$ are zero 
or one  \r\Found. This solution can be obtained by a Baxterization of the braiding matrix of $h$ with $v$, but we shall not need here the explicit solution.
For details of the braiding matrix refer to 
\REF\Moore{G.W. Moore and N. Seiberg, ``Polynomial equations for rational conformal field theories",
Phys. Lett. B 212 (1988) 451.}\r\Moore.
Our focus will be on the crossing relation. This is given by
\def\frac#1#2{{#1 \over #2}}
$$
R^{h,\bar{h}} \pmatrix{ d & c \cr a & b  \cr } (u)=\left(\frac{\psi_a \psi_c}{ \psi_b \psi_d}\right)^{1/2}
R^{h,h}\pmatrix { a & d \cr b & c \cr} (l-u), \e$$
where $l$ is the crossing parameter given by
$$
l=\pi \Delta_{\rm adjoint}/2,\e
$$
where $\Delta_{\rm adjoint}$ is the conformal dimension of the adjoint representation
(assuming a WZW model 
\REF\Witten{E. Witten, ``Nonabelian bosonization in two dimensions", Comm. Math. Phys. 92 (1984) 455.}\r\Witten\
or similar, for a review see e.g. \r\Francesco).
Generally, it is the conformal dimension of the lowest dimensional non--unit field in the fusion product of $h$ and $\bar h$.
We denoted by $R^{h,\bar h}$ and $R^{h,h}$ the trigonometric solution of the YBE based on the
braiding of $h$ with $\bar h$, or $h$ with $h$, respectively.

The $\psi_a$ in eq. (2.5) are called the crossing multipliers. These are given, conjecturally, by
\r\Found
$$
\psi_a={S_{a,0}\over S_{0,0}},
\e $$
where $S_{a,b}$ is the matrix of modular transformation for the primary fields $a$ and $b$,
and $0$ denotes the unit primary field. For an explanation of these notions, see, e.g.,
\r\Francesco.

We wish to describe the crossing relation for the elliptic solution of the YBE. Roughly,
the elliptic solution is given by replacing $\sin u$ in the trigonometric solution, with the theta function
$$
\theta_1(u,q)=2|q|^{1\over 8} |\sin u \prod_{n=1}^\infty (1-2 q^{ n} \cos 2u+q^{2 n})(1-q^{ n}),  \e$$
where $q$ is some parameter $-1<q<1$, called the elliptic modulus. We call this, a thermalization of
the IRF model.

Our conjecture is that the thermal crossing relation
remains the same as in eq. (2.5), except that we need to change the crossing multiplier.
It is given by
$$
\psi_a^t=\chi_a(z,(q^\prime)^\alpha)/\chi_0(z,(q^\prime)^\alpha),  
\e $$
where $\chi_a$ is the character in the RCFT $\cal O$ of the primary field $a$, defined as
$$
\chi_a(z,q)=\sum_{{\cal H}_a} q^{L_0-c/24} z^{J_0}, 
\e $$
where ${\cal H}_a$ is the representation with the highest weight $a$ and $L_0$ is the dimension of the
fields in this representation and $c$ is the central charge. 
$J_0$ is some current in the extended algebra of the theory and $z$ is some gradation of the character which will
be specified below in examples.
Since we will be considering ratios of characters,
we can ignore the factor of $c$. We define
$$ q=\exp(-\epsilon), \e $$
and
$$ q^\prime=\exp(-4 \pi^2/\epsilon),  \e$$
i.e., $q^\prime$ is a modular transformation of $q$. The parameter $\alpha$ in eq. (2.9) is some exponent, which we will be specified later. 

We wish to show that in the critical limit $q \rightarrow 0^+$ the thermalized crossing multiplier becomes
the critical crossing multiplier, $\psi_a$, eq. (2.7). In this limit, it is clear that
$q^\prime\rightarrow 1^-$. Then, using a modular transformation,
$$  \chi_a(1)=\sum_b S_{a,b} \chi_b(0)=S_{a,0},  \e $$  
since $\chi_b(0)=\delta_{a,b}$. Thus we find
$$ \lim_{q\rightarrow 0} \psi^t_a(q)=S_{a,0}/S_{0,0}=\psi_a,\e$$ 
which is the desired relation. The gradation of the character, i.e. $z$, does not change this conclusion.

The thermalized crossing relation, eqs. (2.5, 2.9) was established before in explicit IRF models,
such as, the $A_n$ height models of Jimbo et al.
\REF\Jimbo{M. Jimbo, T. Miwa and M. Okado, ``Local state probabilities of solvable lattice models:
an A(1)(n-1) family", Nucl. Phys. B 300 (1988) 74.}
\r\Jimbo,
the $BCD$ height models 
\REF\BCD{M. Jimbo, T. Miwa and M. Okado, ``Solvable lattice models related to the vector representation of classical
simple Lie algebras", Com. Math. Phys. 116 (1988) 507.}\r\BCD\
and $G_2$ models by Kuniba et al.
\REF\Kuniba{A. Kuniba and J. Suzuki, ``Exactly solvable G2(1) solid on solid model",
Tokyo U., Komaba (1991).}
\REF\Kunibatwo{A. Kuniba, ``Quantum R matrix for G(2) and a solvable 173 vertex model", J. Phys. A
23 (1990) 1349.}
\r{\Kuniba,\Kunibatwo}.

These models correspond in our language to IRF$({\cal O},h,h)$ where the RCFT 
$\cal O$ is a WZW model based on the algebras $A_n$, $B_n$, $C_n$, and $G_2$ 
and the primary field $h$ is the fundamental for $A_n$, the vector
for $B_n,C_n,D_n$ and the $7$ representation for $G_2$.
In all these examples the crossing multiplier can be summarized neatly by the formula
\r{\Jimbo,\BCD,\Kuniba,\Kunibatwo}
$$
\psi_a^t(q)=C \prod_{\alpha\in{\Delta}_+} \theta_1\left({\pi ({\lambda}_a+{\rho}, \alpha)\over k+g},q\right),  \e
$$
where ${\lambda}_a$ is the finite part of the highest weight of the representation $a$ (in the next sections the subscript $a$ will be omitted), ${\rho}$ is half the sum of finite positive roots (also known as the finite counterpart of the Weyl vector $\rho$), ${\Delta}_+$ are the finite positive roots of the algebra and the product $(.,.)$ denotes the product of two weights in the sense of the bilinear form. $k$ is the level of the WZW model and $g$ is
the dual Coxeter number. $C$ is an irrelevant constant.

Actually, the crossing formula, eq. (2.15), is known in the literature to be given by the principally specialized
character \r\BCD, 

$$
\psi_a^t(q)=\prod_{\alpha\in{\Delta}_+} \theta_1\left({\pi (\lambda_a+{\rho},\alpha)\over k+g},q\right)=C\, \chi_a(z,(q^\prime)^{g/(k+g)}), 
\e$$
where $\chi_a(z,q)$, eq. (2.10),  is the character  of the affine algebra $\hat G$ with the highest weight $\lambda_a$. $C$ is  an irrelevant constant that does not depend on $\lambda_a$. 
Also, 
$$z_i=(q^\prime)^{\alpha_i\cdot\rho/(k+g)},\e$$
which for simply laced algebras is the principal gradation. 
Thus, we see that for WZW models eq. (2.9) holds with the exponent  $\alpha=g/(k+g)$. Here, $q$ and $q^\prime$ are given by eqs. (2.11, 2.12). We summarize a proof of eq. (2.16) in the appendix,
since it was not explicit in the literature.

In the critical limit, $q\rightarrow0$ the relation eqs. (2.14, 2.15) becomes
$$
{S_{a,0}\over S_{0,0}}=\prod_{\alpha\in{\Delta}_+} {\sin[\pi ({\lambda}_a+{\rho}, \alpha)/(k+g)]\over
\sin[\pi ({\rho},\alpha)/(k+g)]},\e$$
which is a known formula. For a proof see, e.g., 
\REF\Kac{V.G. Kac, ``Infinite dimensional Lie algebra", Cambridge Univ. Press, 3 ed., 1990}\r\Kac.

\mysec{Local state probability}

We calculate the local state probability following Baxter's corner transfer matrices method (CBM),
described in ref. \r\Baxter\ chapters 13 and 14. For this purpose we first need to use the two inversion
relations.

We define $R_i(u)$ as in eq. (2.3). As before we assume some conformal field theory $\cal O$ and some
primary field in it $h$, not necessarily real. As before we denote by $R^{h,h}_i(u)$ the solution for braiding
$h$ with $h$ and similarly $R^{h,\bar h}_i(u)$ the solution for braiding of $h$ with $\bar h$.
The Yang Baxter equation (YBE) then assumes the form,
$$R_i^{h,\bar h} (u) R^{h,\bar h}_{i+1}(u+v) R^{h,h}_i(v)=R_{i+1}^{h,h}(v) R_i^{h,\bar h}(u+v) R^{h,\bar h}_{i+1}(u),\e$$
along with the relation eq. (2.4) for $R^{h,h}(u)$.

As in ref. \r\Free, we conjecture the first inversion relation to be
$$R_i^{h,h}(u) R_i^{h,h}(-u)=\rho(u)
 \rho(-u) 1_i,\e$$ 
where 
$$\rho(u)=\prod_{r=0}^{n-2} \theta_1(\zeta_r-u,q)/\theta_1(\zeta_r,q),\e$$
where  the theta function was defined in eq. (2.8) 
and
$$\zeta_r={\pi\over 2} (\Delta_{r+1}-\Delta_r).\e$$
Here $\Delta_r$ is the dimension of the field $\psi_r$ appearing in the fusion product,
$$h\cdot h=\sum_{r=0}^{n-1} \psi_r,\e$$
and $r=0,1,\ldots,n-2$.

For the second inversion relation we need the fusion product
$$h \cdot \bar h=\sum_{r=0}^{n-1} \twidle \psi_r,\e$$
We denote by $\twidle\Delta_r$ the dimension of $\twidle\psi_r$.
Of particular significance is the dimension of
$\twidle\Delta_1$,
which is the dimension of the adjoint representation in WZW model, so we denote it as
$\Delta_{\rm adjoint}$, in general. For WZW models we have,
$$\Delta_{\rm adjoint}={g\over  k+g},\e$$
where $g$ is the dual Coxeter number and $k$ is the level
\REF\KZ{ V.G. Knizhnik and A.B. Zamolodchikov, ``Current algebra and Wess--Zumino model in two dimensions", Nucl. Phys. B 247 (1984) 83.}
\r\KZ.
The crossing parameter is 
$$l=\pi\Delta_{\rm adjoint}/2.\e$$

The crossing relation, eq. (2.5), relates $R^{h,h}_i(u)$ with $R_i^{h,\bar h}(u)$.
The crossing parameter is $l$ and the crossing multipliers are given by eq. (2.9),
where $\alpha$ can be seen to be given by,
$$\alpha=\Delta_{\rm adjoint},\e$$
in WZW models where $\Delta_{\rm adjoint}$ is given by eq. (3.7). We conjecture that this is true in general
and $\alpha$ is always given by $\Delta_{\rm adjoint}=\twidle \Delta_1$.

The second inversion relation is then seen to be,
$$R_i^{h,\bar h}(u) R^{h,\bar h}_i(-u)=\twidle\rho(u) \twidle\rho(-u) 1_i,\e$$
where
$$\twidle \rho(u)=\prod_{r=0}^{n-2} \theta_1(\twidle \zeta_r-u,q)/\theta_1(\zeta_r,q),\e$$
where
$$\twidle\zeta_r={\pi\over2} (\twidle\Delta_{r+1}-\twidle\Delta_r),\e$$
and $r=0,1,\ldots,n-2$.

We can now turn to the calculation of the local state probability. 
This is the probability to find at the origin of the lattice some primary field, given some boundary
conditions (which are taken to be in the ground state values).
Since we have the two inversion relations,
eqs. (3.2,  3.10), we can invoke Baxter's corner transfer matrix method \r\Baxter, chapters 13,14 and
appendix A of 
\REF\ABF{G.E. Andrews, R.J. Baxter and P.J. Forrester, ``Eight vertex SOS model and generalized Rogers--Ramanujan type identities", J. Stat. Phys. 35 (1984) 193.} 
\r\ABF. 
We focus on regime III
of the model which is defined by
$$0<q<1,\qquad {\rm and\ } 0<u<l.\e$$
We denote,
$$q=e^{-\epsilon},\quad x=\exp[-4\pi^2/\epsilon],\quad w=\exp[-4\pi u/\epsilon].\e$$
We define the matrix
$$A_{\sigma,\sigma^\prime}(u),\e$$
as the south--east corner transfer matrix for a lattice size $m$ and $\sigma$ stands for the heights along
the negative $y$--axis, $\sigma=\{\sigma_1,\sigma_2,\ldots,\sigma_m\}$. Similarly,
$\sigma^\prime$ stands for the spins along the positive $x$-axis. We take on the edges the ground state
values for the spins.
We define the diagonal matrix $S$ as,
$$(S_{a})_{\sigma,\sigma^\prime}=\delta_{\sigma_1,a}  \prod_{i=1}^m \delta_{\sigma_i,\sigma_i^\prime},\e$$
and the diagonal matrix  $\cal G$ as,
$${\cal G}_{\sigma,\sigma^\prime}= \psi_{{\sigma_1}}^t \prod_{i=1}^m \delta_{\sigma_i,\sigma_i^\prime},\e$$

The local state probability, which is the probability of finding the state $\sigma_1$ at the origin,
is then seen to be given by \r\Baxter,
$$ P(a)=\Tr[{\cal G} S_a A( 2 l)]/\Tr[{\cal G}  A(2 l)].\e$$

The diagonal form of the matrix $A$, denoted by $A_d$, is given by
$$(A_d)_{j,j}=\exp( \alpha_j u),\e$$
where $j=1,2,\ldots$, labels the eigenvalues and we used the first inversion relation, eq. (3.2). 
Now, we assume that the Boltzmann weights are
functions of $w$, eq. (3.14). Then, taking $u\rarrow u+\half i\epsilon$ does not change $w$.
So, $\alpha_j=-4\pi n_j/\epsilon$,
and
$$(A_d)_{j,j}=w^{n_j},\e$$
where $n_j$ are integers.
We can calculate the integers $n_j$ choosing any limit we wish for $u$, since these integers cannot change continuously
as functions of $u$. We choose the limit,
$$x\rarrow 0^+,\quad w={\rm fixed}.\e$$
In this limit, the matrix $A$ usually become diagonal in regime III and we can easily calculate $n_j$.
The Boltzmann weights are then, in this limit,
$$\omega\pmatrix{d&c \cr a&b\cr}=\delta_{a,c} w^{H(a,b,d)},\e$$
where $H$ is some integer function.
The diagonal transfer matrix then becomes,
$$(A_d)_{\sigma,\sigma}=w^{n(\sigma)},\e$$

where 
$$n(\sigma)=\sum_{i=1}^m i[ H(\sigma_{i+1},\sigma_{i+2},\sigma_i)-
H(\bar\sigma_{i+1},\bar\sigma_{i+2},\bar\sigma_i)]
,\e$$
and $\bar\sigma_i$ is the ground state value of $\sigma_i$.
Substituting in eq. (3.18) we find an expression for the local state probability,
$$P(a)=\sum_{\sigma} {\cal G}  S_a w_0^{2 n(\sigma)}\big/\sum_{\sigma} {\cal G} w_0^{2 n(\sigma)},\e$$
where $w_0$ is the value of $w$ at $u=l$,
$$w_0^2=  \exp[-4\pi^2\Delta_{\rm adjoint}/\epsilon]=x^{\Delta_{\rm adjoint}}.\e$$

The expression
$$N(a,b,c)=\sum_\sigma S_a w_0^{2n(\sigma)},\e$$
is known as the one dimensional configuration sum. 
Here $b$ and $c$ are some primary fields labeling the ground states in regime III.
In many models, based on WZW CFT, \REF\models{E. Date, M. Jimbo and M. Okado,  ``Crystal base and Q vertex operators", Comm. Math. Phys. 155 (1993) 47.}
\REF\One{E. Date, M. Jimbo, A. Kuniba, T. Miwa and M. Okado, ``One--dimensional configuration sums
in vertex models and affine Lie algebras characters", Lett. Math. Phys. 17 (1989) 69.}
\r{\Jimbo,\BCD,\models,\One}, the one dimensional configuration sum is seen to be given by the branching function of some coset model, which is the fixed
point conformal field theory, in regime III. In ref. \r\Free, it was conjectured that the fixed point CFT is
given in regime III by the coset, 
$$G/{\cal O},\e$$
where $\cal O$ is the original CFT used to define the model and $G$ is some CFT.
We find it convenient to assume that
$G$ is a product of two CFT $G=G_1\times G_2$. We than see that the one dimensional 
configuration sum is
$$N(a,b,c)=B_{a,b,c}(x^{\Delta_{\rm adjoint}}),\e$$
where the branching function of the coset is defined by
$$\chi_b(z_1,q) \chi_c(z_2,q)=\sum_a \chi_a(z,q) B_{a,b,c}(q),\e$$
where $z$ is a vector indicating any gradation and
$$q=x^{\Delta_{\rm adjoint}}.\e$$
Here $z_1$ and $z_2$ denote the gradations according to the coset rule. In many models $z=z_1=z_2$.
Here $b$ is a primary field of the CFT $G_1$ and $c$ that of $G_2$.
Based on the examples in the literature \r{\Jimbo,\BCD,\One,\models}\ we conjecture that the one dimensional configuration sum 
is always given by the branching function, eqs. (3.29, 3.30). We shall assume this conjecture.

From eq. (2.9) we find that the crossing parameter is given by
$$\psi_a(z,(q^\prime)^\alpha),\e$$
and so it is, from eq. (3.9),
$$\psi_a(z,x^{\Delta_{\rm adjoint}}),\e$$
where $z$ is the gradation vector.

Substituting the one dimensional configuration sum into eq. (3.25), we find for the local state
probability,
$$P(a | b,c)={\chi_a(z,q) B_{a,b,c}(q)\over \chi_b(z_1,q) \chi_c(z_2,q)},\e$$
where we used eqs. (3.29) and
$q=x^{\Delta_{\rm adjoint}}$,
We used the branching rule eq. (3.30)
to show that
$$\sum_a P(a | b,c)=1,\e$$
as it should be.
Here $z$, $z_1$ and $z_2$,  are the gradations of the algebra and it is given, for WZW models by
$$z_i=x^{\rho\cdot \alpha_i/(k+g)},\e$$
and $z=z_1=z_2$.
Here $\rho$ is half the sum of positive roots and $\alpha_i$ is the $i$th simple root.

\mysec{Discussion.}

In this paper we discussed the local state probabilities for a large class of lattice models,
constructed from an arbitrary conformal field theory \r\Found. This result complements the calculation of the free energies of such models \r\Free\  and based on that it was conjectured
that in regime III of the models, the critical CFT is given by $G/{\cal O}$, where $\cal O$ is the CFT used to
build the model. Here we find support for this conjecture by showing that it is consistent with 
the general expression that we find for the local state probabilities.

The main examples for such models are WZW theories, see, e.g., \r{\Jimbo,\BCD,\models}. 
Other models based on other CFT's are known in the literature, see 
\REF\Baver{E. Baver, ``New solvable lattice models from conformal field theory", Phys. Lett. B 387 (1996) 502.}\r\Baver, but for the most part these models remain to be explored explicitly in the future.

We propose,
based on our work,
that in the scaling regime all these lattice models correspond to integrable field theories. It is intriguing
to try to evaluate the factorized scattering matrices of such theories.

\appendix 
\line{\bf The crossing multiplier as a character.\hfill}

\par
We wish to calculate the principally graded character at nome
$$(q^\prime)^{g/(k+g)},\e$$
where $q=e^{-\epsilon}$ and $q^\prime=\exp(-4\pi^2/\epsilon)$.
The principally graded character is given by (\r\Kac, eq. (10.9.4) there)
$$\chi=p^{-\Lambda\cdot \rho} \prod_{\alpha\in\Delta_+} (1-p^{(\Lambda+\rho)\cdot\alpha})^{{\rm mult}\,\alpha},\e$$
up to an irrelevant factor which does not depend on $\Lambda$. Here $\Delta_+$ are the positive roots,
$\Lambda$ is the highest weight and $\rho$ is the affine half sum of positive roots. The algebra is
affine $\hat G$ and the finite algebra is denoted by $G$.
Now since $q^\prime$ counts $L_0$, eq. (2.10), which is $\rho.\delta=g$ simple roots we have
$$p=(q^\prime)^{1/(k+g)}.\e$$
The positive roots of the affine algebra are
$m\delta$ at multiplicity $n=rank \,G$, and
$$(m-1)\delta+\bar\alpha, \qquad m\delta-\bar\alpha,\e$$
where $m=1,2,\ldots$ and $\bar\alpha$ is any positive root of the finite algebra $G$ and $\delta$ is 
the simple imaginary root. For explanation of these notions see \r\Kac.
Thus, eq. (A.2) becomes,
$$\chi=p^{-\Lambda\cdot\rho}\prod_{m=1}^\infty \prod_{\bar\alpha\in\bar\Delta_+} (1-x^{m-1} w_{\bar\alpha})(1-x^m w_{\bar\alpha}^{-1})
(1-x^m),\e$$
up to an irrelevant factor which does not depend on $\Lambda$. Here
$$x=p^{k+g}=q^\prime,\e$$
and
$$w_{\bar \alpha}=(q^\prime)^{(\bar\Lambda+\bar\rho)\cdot\bar\alpha/(k+g)}.\e$$
We now use Baxter's formula (\r\Baxter, eq. (14.2.42) there),
$$\theta_1(u,e^{-\epsilon})=\exp[2u(\pi-u)/\epsilon] f(e^{-4\pi u/\epsilon},e^{-4\pi^2/\epsilon}),\e$$
up to an irrelevant factor and where
$$f(w,y)=\prod_{n=1}^\infty (1-y^{n-1} w)(1-y^n w^{-1})(1-y^n).\e$$
Since $q=e^{-\epsilon}$ and $q^\prime=e^{-4\pi^2/\epsilon}$ we find
$$\chi=\prod_{\bar\alpha\in\bar\Delta_+} \theta_1(\pi(\bar\Lambda+\bar\rho)\cdot\bar\alpha/(k+g),q),\e$$
which is the desired formula, eq. (2.16). 
The factor in eq. (A.8) cancels exactly the factor of
$$p^{-\Lambda\cdot \rho},\e$$
in eq. (A.2) and thus this formula, eq. (A.10), is correct. To see this define
$$u_\alpha=\pi(\bar\Lambda+\bar\rho)\cdot\bar\alpha/(k+g).\e$$
Remember also that 
$$\Lambda=\bar\Lambda-m_\Lambda \delta+k\Lambda_0,\e$$
where 
$$m_\Lambda={(\bar\Lambda+\bar\rho)^2\over 2(k+g)},\e$$
up to a summand that does not depend on $\bar\Lambda$.
Now, we have the equation
$$p^{-\bar\Lambda\cdot \bar\rho} p^{m_\Lambda g} \prod_{\alpha\in\bar\Delta_+}
\exp[2 u_\alpha(\pi-u_\alpha)/\epsilon]=1,\e$$
which shows that the factors cancel exactly, and thus, eq. (A.10), is correct, up to factors that do not depend
on $\Lambda$.

\ack

We thank Paul Pearce and Michio Jimbo for many fruitful discussions and Ida Deichaite for remarks
on the manuscript.

\refout

\bye